\documentclass{emulateapj}

\shorttitle{Electron-capture supernovae as the origin of elements beyond
iron}

\shortauthors{Wanajo et al.}

\begin{document}


\title{Electron-capture supernovae as the origin of elements beyond iron}

\author{Shinya Wanajo\altaffilmark{1, 2},
        Hans-Thomas Janka\altaffilmark{2},
        and 
        Bernhard M\"uller\altaffilmark{2}
        }

\altaffiltext{1}{Technische Universit\"at M\"unchen,
        Excellence Cluster Universe,
        Boltzmannstr. 2, D-85748 Garching, Germany;
        shinya.wanajo@universe-cluster.de}

\altaffiltext{2}{Max-Planck-Institut f\"ur Astrophysik,
        Karl-Schwarzschild-Str. 1, D-85748 Garching, Germany;
        thj@mpa-garching.mpg.de}

\begin{abstract}
  We examine electron-capture supernovae (ECSNe) as sources of
  elements heavier than iron in the solar system and in Galactic halo
  stars. Nucleosynthesis calculations are performed on the basis of
  thermodynamic histories of mass elements from a fully
  self-consistent, two-dimensional (2D) hydrodynamic explosion model
  of an ECSN. We find that neutron-rich convective lumps with an
  electron fraction down to $Y_\mathrm{e, min} = 0.40$, which are
  absent in the one-dimensional (1D) counterpart, allow for
  interesting production of elements between the iron group and $N=50$
  nuclei (from Zn to Zr, with little Ga) in nuclear
  (quasi-)equilibrium.  Our models yield very good agreement with the
  Ge, Sr, Y, and Zr abundances of r-process deficient Galactic halo
  stars and constrain the occurrence of ECSNe to $\sim$4\% of all
  stellar core-collapse events.  If tiny amounts of additional
  material with slightly lower $Y_\mathrm{e, min}$ down to
  $\sim$0.30--0.35 were also ejected ---which presently cannot be
  excluded because of the limitations of resolution and
  two-dimensionality of the model---, a weak r-process can yield
  elements beyond $N=50$ up to Pd, Ag, and Cd as observed in the
  r-process deficient stars.
\end{abstract}


\section{Introduction}
Electron-capture supernovae (ECSNe) have been proposed as
possible origin of elements beyond iron, in particular of
heavy r-process elements. This is due to the distinctive features of
the collapsing O-Ne-Mg cores \citep{Nomo1987}. The cores with their fairly
small mass ($\sim$1.38$\,M_\odot$), steep surface density gradient,
and dilute H/He envelope (in the super-asymptotic-giant-branch or
SAGB stage) were expected to enable prompt explosions, where the ejection
of neutron(n)-rich matter with an electron fraction (number of protons per 
baryon) of $Y_\mathrm{e} \approx 0.14$ could be possible \citep{Wana2003}.
Chemical evolution studies also favor ECSNe at
the lower end ($\sim$8--10$\,M_\odot$) of the mass range of
core-collapse supernovae (CCSNe) to account for
the abundance signatures of Galactic halo stars \citep{Ishi1999}.

Recent one-dimensional (1D) hydrodynamic simulations of O-Ne-Mg core
collapse with elaborate neutrino transport, however, found 
delayed neutrino-driven explosions \citep{Kita2006}. This result 
was confirmed by \citet{Dess2006} and excludes the prompt explosions
of \citet{Hill1984}. The shocked surface layers of the
O-Ne-Mg core are also far from the r-process conditions envisioned 
by \citet{Ning2007} as shown by \citet{Jank2008}.

Detailed nucleosynthesis studies based on the 1D results of
\citet{Kita2006}, however, suggest that ECSNe can be the source of
nuclei up to $N=50$, in particular of Zn and light p-nuclei (up to
$^{92}$Mo), whose origins are not fully understood (Wanajo et
al. 2009; see also Hoffman et al. 2008). \citet{Wana2009} further
speculated that the nucleosynthetic outcome could be considerably
altered in the multi-dimensional case.

Spectroscopic studies of Galactic halo stars indicate the
presence of a significant fraction of r-process \textit{deficient}
Galactic halo stars that possess high Sr-Y-Zr ($N=50$) abundances
relative to heavy r-process material \citep{John2002, Hond2006}.
This implies the contribution from an astrophysical source making
$N=50$ species without heavy r-process elements
\citep{Trav2004, Wana2006, Qian2008}. Recently \citet{Arco2010},
employing trajectories from 1D hydrodynamic models and varying
the corresponding $Y_\mathrm{e}$, proposed that the
baryonic wind blown off the surface of new-born (``proto-'') neutron 
stars (PNSs) by neutrino heating could be production sites of
such nuclei in the early Galaxy.

New 2D explosion simulations of ECSNe exhibit n-rich lumps of
matter being dredged up by convective overturn from the outer layers
of the PNS during the early stages of the explosion (\S~2; M\"uller,
Janka, \& Kitaura, in preparation), a feature that is absent in the 1D 
situation.
In this \textit{Letter} we present the first nucleosynthesis study
on the basis of such self-consistent 2D models. Without invoking
any free parameters we show that ECSNe could be an important
source of the elements between the iron peak and the
$N=50$ region in the Sun and the early Galaxy, perhaps even of 
light r-process nuclei up to Cd.

\section{Two-dimensional Supernova Model}

\begin{figure}
\epsscale{1.0}
\plotone{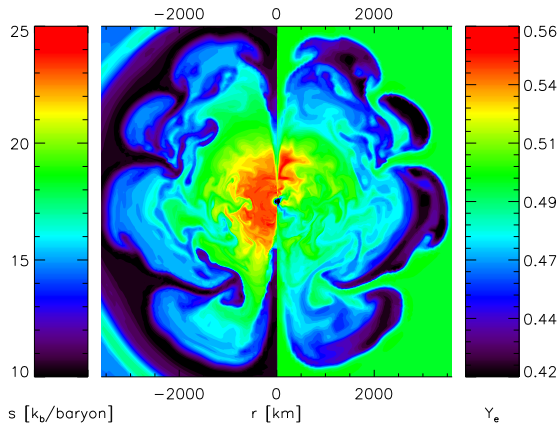}
\caption{Snapshot of the convective region of the 2D simulation of 
an ECSN at 262~ms after core bounce with entropy per nucleon ($s$; left) 
and $Y_\mathrm{e}$ (right). Mushroom-shaped lumps of low-$Y_\mathrm{e}$
matter are ejected during the early phase of the explosion.
}
\label{fig:2Dsnapshot}
\end{figure}

\begin{figure}
\epsscale{1.0}
\plotone{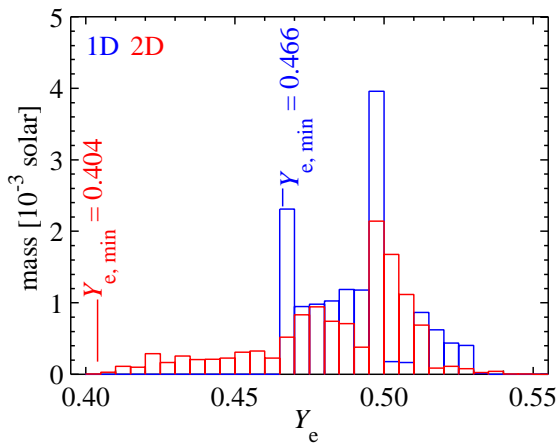}
\caption{Ejecta masses vs.\ $Y_\mathrm{e}$ for the
1D (blue) and 2D (red) explosion models. The width of a $Y_\mathrm{e}$-bin 
is chosen to be $\Delta Y_\mathrm{e} = 0.005$. The minimum values of
$Y_\mathrm{e}$ are indicated for both cases.
}
\label{fig:massvsYe}
\end{figure}

The nucleosynthesis analysis made use of about 2000 representative
tracer particles, by which the thermodynamic histories of ejecta
chunks were followed in our 2D hydrodynamic calculation of an
ECSN. The model was computed with a sophisticated (ray-by-ray-plus)
treatment of the energy-dependent neutrino transport, using the
\textsc{Prometheus-Vertex} code and the same microphysics
\citep[weak-interaction rates, nuclear burning treatment, and nuclear
equation of state of][]{Latt1991} as in its 1D counterpart
\citep{Kita2006}. Some aspects of the 2D model in comparison to 1D
results were discussed by \citet{Jank2008}, and a more complete
description can be found in a separate publication (M\"uller et al.,
in preparation).

The pre-collapse model of the O-Ne-Mg core emerged
from the evolution of an 8.8$\,M_\odot$ star \citep{Nomo1987}.
Because of the very steep density gradient near the core
surface, the shock expands continuously, and a neutrino-powered
explosion sets in at $t \sim 100\,$ms p.b. in 1D and 2D essentially 
in the same way and with a very similar energy 
\citep[$\sim$10$^{50}\,$erg;][]{Jank2008}.

In the multi-dimensional case, however, the negative entropy profile 
created by neutrino heating around the PNS leads to a short
phase of convective overturn, in which accretion downflows deleptonize
strongly, are neutrino heated near the neutrinosphere, and rise 
again quickly, accelerated by buoyancy forces. 
Thus n-rich matter with modest entropies per nucleon 
($s \sim$13--15$\,k_\mathrm{B}$; $k_\mathrm{B}$ is Boltzmann's constant)
gets ejected in 
mushroom-shaped structures typical of Rayleigh-Taylor instability.
Figure~\ref{fig:2Dsnapshot} displays 
the situation 262$\,$ms after bounce when the 
pattern is frozen in and self-similarly expanding. 

As a consequence, the mass distribution of the ejecta in the 2D model
extends down to $Y_\mathrm{e, min}$ as low as $\sim$0.4,
which is significantly more n-rich than in the corresponding 1D case
($Y_\mathrm{e, min}^\mathrm{1D}\sim 0.47$)\footnote{Note that the 
exact lower bound of the mass distribution vs.\ $Y_\mathrm{e}$ in the 
1D case is highly sensitive to details of the neutrino transport,
e.g.\ the number and interpolation of grid points in energy space. 
In a recent simulation with improved spectral resolution, \citet{Hued2010}
obtained $Y_\mathrm{e, min} = 0.487$.}. Figure~\ref{fig:massvsYe} 
shows the $Y_\mathrm{e}$-histograms at the end of the simulations. 
The total ejecta masses are $1.39\times 10^{-2}\,M_\odot$ for the 1D model
and $1.14\times 10^{-2}\,M_\odot$ in 2D, where the difference is
partly due to the different simulation times, being $\sim$800$\,$ms and
$\sim$400$\,$ms, respectively (core bounce occurs at $\sim$50$\,$ms). 
However, the ejecta after $\sim$250$\,$ms p.b.\ are only proton-rich,
contributing merely to the $Y_\mathrm{e} > 0.5$ side in 
Fig.~\ref{fig:massvsYe}.

\section{Nucleosynthesis for the ECSN Model}

\begin{figure}
\epsscale{1.0}
\plotone{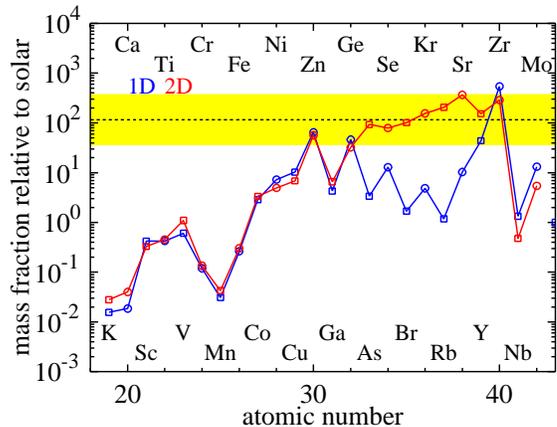}
\caption{Elemental mass fractions in the ECSN ejecta relative to their
solar values \citep{Lodd2003}, comparing the 2D results (red) 
with the 1D counterpart (blue) from \citet{Wana2009}. Even-$Z$ and odd-$Z$
elements are denoted by circles and squares, respectively. The 
normalization band (see text) is marked in yellow.
}
\label{fig:ECSNyields}
\end{figure}

\begin{figure}
\epsscale{1.0}
\plotone{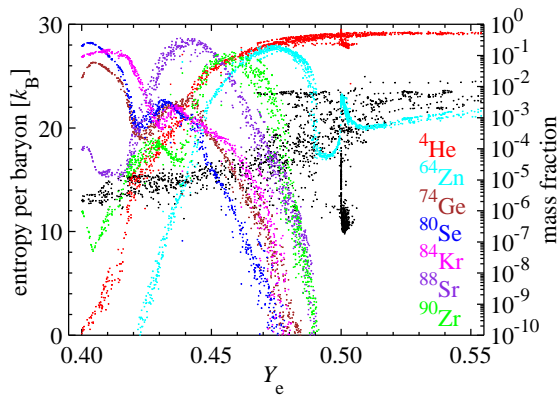}
\caption{Mass fractions of the dominant isotopes of the even-$Z$
elements vs.\ $Y_\mathrm{e}$ 
for all investigated thermodynamic trajectories. Also shown are the
mass fractions of $^4$He ($\alpha$-particles)  (red) and the
entropies of the trajectories (black dots, left vertical axis).
}
\label{fig:massfractions}
\end{figure}

The nucleosynthetic yields are obtained with the reaction network
code (including neutrino interactions) described in \citet{Wana2009}. 
Using thermodynamic trajectories 
directly from the 2D ECSN model, the calculations are started when
the temperature decreases to $9 \times 10^9$~K, assuming initially
free protons and neutrons with mass fractions
$Y_\mathrm{e}$ and $1-Y_\mathrm{e}$, respectively.
The final abundances for all isotopes are obtained by
mass-integration over all 2000 marker particles.

The resulting \textit{elemental} mass fractions relative to solar
values \citep{Lodd2003}, or the production factors, are shown in
Fig.~\ref{fig:ECSNyields} (red) compared to the 1D case (blue)
from \citet{Wana2009}. The ``normalization band''
between the maximum (367 for Sr) and a tenth of that is indicated
in yellow with the medium marked by a dotted line. The total ejecta
mass is taken to be the sum of the ejected mass from the core and the
outer H/He-envelope ($= 8.8\,M_\odot - 1.38\,M_\odot + 0.0114\,M_\odot= 
7.43\,M_\odot$). Note that the $N=50$ species, $^{86}$Kr, $^{87}$Rb,
$^{88}$Sr, and $^{90}$Zr, have the largest production factors for
\textit{isotopes} with values of 610, 414, 442, and 564, respectively.

As discussed by \citet{Wana2009}, in the 1D case only Zn and Zr are on
the normalization band, although some light p-nuclei (up to $^{92}$Mo)
can be sizably produced. In contrast, we find that all elements
between Zn and Zr, except for Ga, fall into this band in the 2D case 
(Ge is marginal), although all others are almost equally produced in
1D and 2D. This suggests ECSNe to be likely sources of 
Zn, Ge, As, Se, Br, Kr, Rb, Sr, Y, and Zr, in the Galaxy. Note
that the origin of these elements is not fully understood, although
Sr, Y, and Zr in the solar system are considered to be dominantly made
by the s-process. The ejected masses of $^{56}$Ni ($\to ^{56}$Fe;
$3.0 \times 10^{-3}\,M_\odot$) and all Fe ($3.1 \times 10^{-3}\,M_\odot$)
are the same as in the 1D case 
\citep[$2.5 \times 10^{-3}\,M_\odot$,][]{Wana2009}.

The fact that oxygen is absent in ECSN ejecta but a dominant product
of more massive CCSNe, can pose a constraint on the frequency of ECSNe
\citep{Wana2009}. Considering the isotope $^{86}$Kr with its largest
production factor in our 2D model and assuming $f$ to be the fraction 
of ECSNe relative to all CCSNe, one gets
\begin{equation}
\frac{f}{1-f} 
= \frac{X_\odot(^{86}\mathrm{Kr})/X_\odot(^{16}\mathrm{O})}
       {M(^{86}\mathrm{Kr})/M_\mathrm{noEC}(^{16}\mathrm{O})}
= 0.050,
\end{equation}
where $X_\odot(^{86}\mathrm{Kr}) = 2.4 \times 10^{-8}$ and
$X_\odot(^{16}\mathrm{O}) = 6.6 \times 10^{-3}$ are the mass fractions
in the solar system \citep{Lodd2003}, $M(^{86}\mathrm{Kr}) = 1.1 \times
10^{-4}\,M_\odot$ is our ejecta mass of $^{86}$Kr, and 
$M_\mathrm{noEC}(^{16}\mathrm{O}) = 1.5\,M_\odot$ the production of 
$^{16}$O by all other CCSNe, averaged over the stellar initial mass
function between $13\,M_\odot$ and $40\,M_\odot$ \citep[see][]{Wana2009, 
Nomo2006}. Equation~(1) leads to $f = 0.048$. The frequency of
ECSNe relative to all CCSNe is thus $\sim$4\%, assuming
that all $^{86}$Kr in the solar system except for a possible
contribution from the s-process \citep[18\%,][]{Arla1999}, originates
from ECSNe. This is in good agreement with the prediction from a recent
synthetic model of SAGB stars \citep[for solar
metallicity models,][]{Poel2008}.

\begin{figure}
\epsscale{1.0}
\plotone{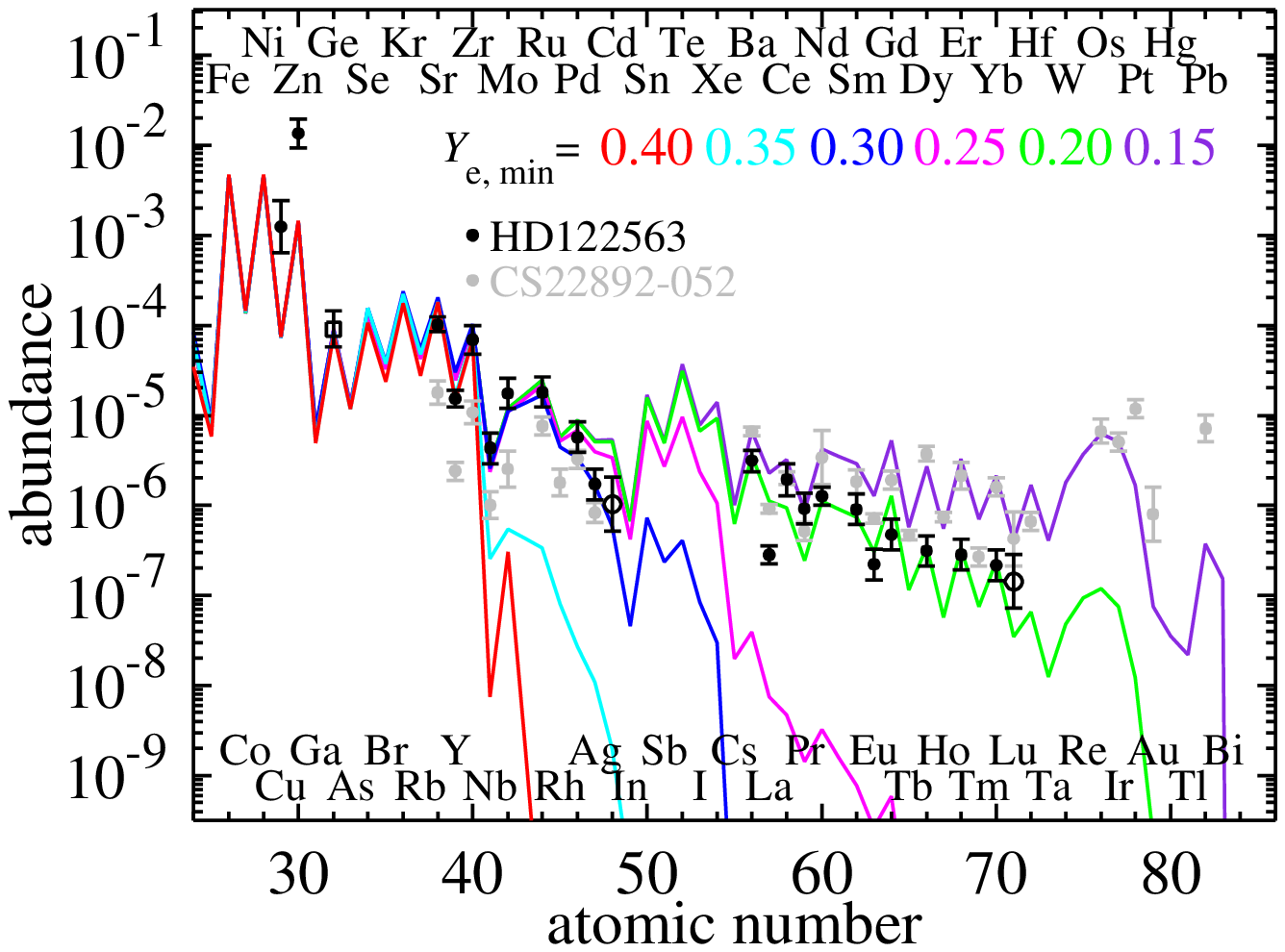}
\caption{Elemental abundances for various
 $Y_\mathrm{e, min}$ compared with the stellar abundances of
 the r-process deficient star HD~122563 with
 [Fe/H]$\,\approx\,-$2.7 \citep{Hond2006} and the r-process
 enhanced star CS~22892-052 with [Fe/H]$\,\approx\,-$3.1
 \citep{Sned2003}. The abundances of HD~122563 and CS~22892-052 are
 vertically shifted to match Zr for $Y_\mathrm{e, min} =
 0.40$ and Ba for $Y_\mathrm{e, min} = 0.15$, respectively.
 The Cd and Lu values in HD~122563 (open circles) are from 
 \citet{Roed2010}, and the Ge value (open square) is from
 \citet{Cowa2005}.
}
\label{fig:weakrprocess}
\end{figure}

The remarkable difference between the 1D and 2D cases
(Fig.~\ref{fig:ECSNyields}) can be understood by the element formation
in nuclear (quasi-)eqilibrium. Representative for all elements in the
normalization band of Fig.~\ref{fig:ECSNyields}, the final mass
fractions of the isotopes $^{64}$Zn, $^{74}$Ge, $^{80}$Se, $^{84}$Kr,
$^{88}$Sr, and $^{90}$Zr \citep[accounting for dominant fractions of
49\%, 36\%, 50\%, 57\%, 82\%, and 51\%, respectively, of their
elements in the solar system,][]{Lodd2003} are displayed for all
tracer trajectories in Fig.~\ref{fig:massfractions}.

Nuclear quasi-equilibrium (QSE) makes nuclei heavier than the Fe-group
up to $A \sim 90$ \citep{Meye1998}. $^{64}$Zn, $^{88}$Sr, and
$^{90}$Zr are thus produced at $Y_\mathrm{e} = 0.43$--0.49, where the
$\alpha$-concentration (at the end of the calculations) is
$X(^4\mathrm{He}) = 0.001$--0.1. The QSE with abundant $\alpha$
particles, however, is known to leave a deep trough in the abundance
curve between $A\sim$60 and 90 because of the strong binding at $N =
28$ and 50.  This explains the substantial underproduction of elements
around $Z \sim$33--37 in the 1D case (Fig.~\ref{fig:ECSNyields}, blue
line).

As the $\alpha$-concentration becomes small, QSE asymptotes to nuclear
statistical equilibrium (NSE). Since NSE with neutron excess
($Y_\mathrm{e} \sim 0.4$) leads to nuclei heavier than the Fe-group up
to $A \approx 84$ \citep{Hart1985}, the trough can be filled by
NSE-abundances assembled in the n-rich ejecta lumps. The small
$X(^4{\mathrm{He}})$ ($< 10^{-5}$) at $Y_\mathrm{e} \sim 0.40$--0.42
(Fig.~\ref{fig:massfractions}) is indicative of the consumption of
almost all $\alpha$-particles before freeze-out from NSE. Accordingly,
QSE asymptotes to NSE in this $Y_\mathrm{e}$-range and yields
substantial amounts of $^{74}$Ge, $^{80}$Se, and $^{84}$Kr (made as
$^{74}$Zn, $^{80}$Ge, and $^{84}$Se), nuclei that cannot be created in
$\alpha$-rich QSE. A similar result can be seen in the QSE study by
\citet{Meye1998} with entropies and $Y_\mathrm{e}$-values close to
ours here (see Fig.~14 in their paper).

In the n-rich ejecta lumps NSE (or $\alpha$-deficient QSE) conditions
are established for several reasons. They have smaller entropies ($s
\approx 13$--$15\,k_\mathrm{B}$ per baryon) than the other ejecta
(where $s \approx 15$--$20\,k_\mathrm{B}$ per baryon;
Figs.~\ref{fig:2Dsnapshot},$\,$\ref{fig:massfractions}).  This favors
$\alpha$-particles to disappear when NSE ends as the temperature
drops. In addition, the $\alpha$'s become easily locked up and tightly
bound in nuclei, i.e., their separation energies are large \citep[cf.,
e.g., Fig.~1b in][]{Woos1992}, because nuclei with n-excess do not
readily release $\alpha$'s to move farther away from
$\beta$-stability.

We find no sign of r-processing in the n-rich lumps. Our present
calculations are limited to the first $\la$400$\,$ms after bounce 
and do not include the neutrino-driven PNS wind. The latter, however,
turned out to have proton excess in 1D models of the long-term 
evolution of ECSNe \citep{Hued2010}. It thus makes only p-rich 
isotopes with small production factors (below unity,
Wanajo et al., in prep.) and has no effect on the discussed
results.

\section{ECSN Yields and Galactic Halo Stars}

Our results in \S~3 also imply that ECSNe can be the source of Sr, Y,
and Zr as observed in r-process deficient Galactic halo 
stars (Fig.~\ref{fig:weakrprocess}). A number of
such stars with detailed abundance determinations indicate, however, a
possible link with the elements beyond $N=50$, e.g.\ Pd and Ag
\citep{Hond2006}. 
Our ECSN models cannot account for the production of such elements, 
but in their ejecta a small change of $Y_\mathrm{e}$ can
drastically change the nucleosynthesis \citep{Wana2009}.
Due to limitations of the numerical resolution and the lack of the 
third dimension, or some sensitivity to the nuclear equation of state,
it cannot be excluded that ECSNe also eject tiny amounts of matter 
with $Y_\mathrm{e, min}$ slightly lower than predicted by the 2D
simulation.

We therefore compare the nucleosynthesis for $Y_\mathrm{e, min} = 0.40$
of our ECSN model and for artificially reduced values of
$Y_\mathrm{e, min} = 0.35, 0.30, 0.25, 0.20$, and 0.15 with the
abundance patterns of representative r-process deficient 
\citep[HD~122563,][]{Hond2006} and enhanced
\citep[CS~22892-052,][]{Sned2003} stars (Fig.~\ref{fig:weakrprocess}).
For that we use the thermodynamic trajectory of the lowest 
$Y_\mathrm{e}$ ($= 0.404$) of the original model but apply $Y_\mathrm{e}$
down to 0.15 in steps of $\Delta Y_\mathrm{e} = 0.005$. The ejecta masses
in these additional $Y_\mathrm{e}$-bins are chosen to be constant with
$\Delta M = 2\times 10^{-5}\,M_\odot$ in the cases
$Y_\mathrm{e, min} = 0.35$ and 0.30, and
$\Delta M = 10^{-5}\,M_\odot$ for the other $Y_\mathrm{e, min}$.

Figure~\ref{fig:weakrprocess} shows that
$Y_\mathrm{e} \la 0.35$ is needed to obtain elements beyond
$N=50$. A remarkable agreement with the abundance pattern in
HD~122563 up to Cd ($Z = 48$) can be seen for $Y_\mathrm{e, min} =
0.30$. Such a mild reduction of 
$Y_\mathrm{e, min}$ in the ECSN ejecta is well possible
for the reasons mentioned
above. A reasonable match of the heavier part
beyond $Z = 48$ requires $Y_\mathrm{e, min} \approx 0.20$.
This, however, leads to a poor agreement for Ag and Cd. We therefore
speculate that ECSNe could be the sources of the elements up to Cd
in r-process deficient stars, and the heavier elements are from
a different origin. Moreover, $Y_\mathrm{e, min} = 0.15$ is 
necessary to reproduce the abundance pattern of r-process 
enhanced stars like CS~22892-052. Such a low $Y_\mathrm{e}$ seems 
out of reach and disfavors ECSNe as production sites of
heavy r-process nuclei.

We note that the neutron-capture reactions start from seeds with $A
\sim 80$ formed in nuclear equilibrium (\S~3), not by an
$\alpha$-capture process. We therefore prefer to call the described
process producing the elements beyond $N=50$, presumably up to Cd,
``weak r-process'' \citep{Wana2006,Hond2006} rather than
$\alpha$-process or charged-particle process
\citep{Woos1992,Qian2008}.

\section{Summary and Outlook}

Using ejecta-mass tracers from a self-consistent 2D explosion model we
computed the nucleosynthesis of ECSNe. Because of a brief phase of
convective overturn and the very fast explosion that is characteristic
of O-Ne-Mg cores, n-rich lumps with $Y_\mathrm{e}$ down to 0.4 are
ejected. These allow for a sizable production of the elements from Zn
to Zr in nuclear (quasi-)equilibrium. The model yields Ge, Sr, Y, and
Zr in very good agreement with abundances of r-process deficient
Galactic halo stars. A mild reduction of the minimum $Y_\mathrm{e}$ to
$\sim$0.30--0.35, which cannot be excluded due to limited numerical
resolution and the lack of the third dimension, leads to a weak
r-process up to the silver region (Pd, Ag, and Cd), again well
matching these elements in r-process deficient stars. The formation of
heavy r-process nuclei requires $Y_\mathrm{e}$ to be as low as
$\sim$0.15--0.20 and seems out of reach for our models.

We therefore determine ECSNe as an important source of Zn, Ge, As, 
Se, Br, Kr, Rb, Sr, Y, and Zr in the solar system and the early 
Galaxy. Our models, however, 
underproduce Ga, which suggests the s-process as the origin of this
element. The frequency of ECSNe is thus constrained to $\sim$4\% of
all CCSN events on average over the Galactic history, but could have
been higher at early Galactic epochs, compatible with the commonality
of r-process deficient halo stars.

Future better resolved and in particular 3D models will have to 
elucidate the role of ECSNe as site of the weak r-process.
Also important are new abundance studies of r-process
deficient stars for more complete information on the elements from
Zn to Zr and on weak r-process products between Nb and Cd.

\acknowledgements We thank L.~H\"udepohl and W.~Aoki for useful
discussions. DFG grants EXC153, SFB/TR27, and SFB/TR7, and computing
time at the NIC in J\"ulich, HLRS in Stuttgart, and the RZG in
Garching are acknowledged.

\end{document}